\documentclass[12pt]{iopart}

\expandafter\let\csname equation*\endcsname\relax
\expandafter\let\csname endequation*\endcsname\relax

\usepackage{mathtools}
\usepackage{graphics}
\usepackage{graphicx}
\usepackage{dcolumn}
\usepackage{bm}
\usepackage{dsfont} 
\usepackage{amsmath,amssymb}
\usepackage{verbatimbox}
\usepackage{hyperref}
\usepackage{tabularx}
\usepackage{epsf,epsfig}
\usepackage[normalem]{ulem}
\usepackage[usenames]{color}
\usepackage{multirow}
\usepackage{makecell}
\usepackage{diagbox}
\usepackage[abs]{overpic}
\allowdisplaybreaks
\hypersetup{
    colorlinks=true,
    linkcolor=blue,
    filecolor=magenta,      
    urlcolor=blue,
    citecolor=blue
}
\newcommand{\I}{{\mbox{\tiny I}}}
\newcommand{\ISCO}{{\mbox{\tiny ISCO}}}
\newcommand{\EdGB}{{\mbox{\tiny EdGB}}}
\newcommand{\RD}{{\mbox{\tiny RD}}}

\newcommand{\GR}{{\mbox{\tiny GR}}}

\newcommand{\MR}{{\mbox{\tiny MR}}}

\definecolor{red(ncs)}{rgb}{0.77, 0.01, 0.2}

\usepackage{array}
\newcolumntype{C}[1]{>{\centering\let\newline\\\arraybackslash\hspace{0pt}}m{#1}}

\newcolumntype{C}[1]{>{\centering\arraybackslash}m{#1}}

\begin{document}


\title[Probing string-inspired gravity with the IMR consistency tests of GWs]{Probing string-inspired gravity with the inspiral-merger-ringdown consistency tests of gravitational waves}

\author{Zack Carson}
\address{Department of Physics, University of Virginia, Charlottesville, Virginia 22904, USA}
\author{Kent Yagi}
\address{Department of Physics, University of Virginia, Charlottesville, Virginia 22904, USA}

\date{\today}


\begin{abstract}
The extreme-gravity collisions between black holes allow us to probe the underlying theory of gravity.
We apply a predictive forecast of the theory-agnostic inspiral-merger-ringdown consistency test to an example theory beyond general relativity for the first time, for future gravitational wave observations. 
Here we focus on the string-inspired Einstein-dilaton Gauss-Bonnet gravity and modify the inspiral, ringdown, and remnant black hole properties of the gravitational waveform. 
We found that future multiband observations allow us to constrain the theory stronger than current observations by an order of magnitude. 
The formalism developed here can easily be applied to other theories.
\end{abstract}

\maketitle


\section{Introduction}\label{sec:intro}
The historic observation of gravitational waves (GWs) from the merger of two black holes (BHs) by the LIGO and Virgo Collaborations (LVC)~\cite{GW150914} has ushered in the birth of a new era of astrophysics, for the first time probing the \emph{extreme gravity} regime where spacetime is strong, non-linear, and dynamical.
GWs such as these carry with them multitudes of information; not only regarding the sources' astrophysical properties, but also about the underlying theory of gravity driving the process.
However, this first event, as well as the following 10~\cite{GW_Catalogue}, have failed to detect any significant deviations from the predictions of general relativity (GR)~\cite{Abbott_IMRcon}, the prevailing theory of gravity for the past century~\cite{Will_SEP}.
While the current LVC infrastructure~\cite{TheVirgo:2014hva,TheLIGOScientific:2014jea} is a marvel of modern engineering, it may not yet be enough to uncover the elusive traces of a modified theory of gravity.
The next generation of GW detectors~\cite{advancedLIGO,Ap_Voyager_CE,ET,LISA,B-DECIGO,DECIGO,TianQin}, on the other hand, promise improvements on the order of $100$ times the sensitivity, as well as new sensitivity in the mHz regime.
Will this be enough to pull back the curtain on the hidden theories of gravity running the show?

Throughout the last century, countless tests and observations of GR have been performed~\cite{Will_SolarSystemTest,Stairs_BinaryPulsarTest,Wex_BinaryPulsarTest,Ferreira_CosmologyTest,Clifton_CosmologyTest,Joyce_CosmologyTest,Koyama_CosmologyTest,Salvatelli_CosmologyTest,Berti_ModifiedReviewLarge,Abbott_IMRcon,Yunes_ModifiedPhysics}, all finding agreement with Einstein's theory in a variety of environments.
However, even with such success, GR still needs to be tested.
While it explains a majority of our observations, there yet remain several unanswered questions which could be explained by new theories of gravity.
For example, ``dark energy" and the accelerated expansion of the universe~\cite{Jain:2010ka,Salvatelli:2016mgy,Koyama:2015vza,Joyce:2014kja}, ``dark matter"
\footnote{Galactic rotation curves as well as other observations can be well explained by dark matter particle models, as well as certain modified theories of gravity~\cite{Famaey:2011kh,Milgrom:DarkMatter,Milgrom:2008rv,Clifton:2011jh,Joyce:2014kja} , although the former typically gives stronger agreement with various observations.}, and more~\cite{Clifton:2011jh,Famaey:2011kh,Joyce:2014kja,Koyama:2015vza,Milgrom:2008rv,Jain:2010ka} all remain open to this day.
To date, a plethora of modified theories of gravity have been proposed to explain some of the open questions listed above.

A particularly interesting and well-studied class of theories involves the addition of a massless scalar field, known as scalar-tensor theories (STTs).
Specifically, we here focus on Einstein-dilaton Gauss-Bonnet (EdGB) gravity motivated from string theory, where the dilaton scalar field couples linearly to a quadratic curvature term in the action~\cite{Kanti_EdGB,Maeda:2009uy,Sotiriou:2013qea}.
Such a coupling allows for the scalarization of BHs~\cite{Campbell:1991kz,Yunes:2011we,Takahiro,Sotiriou:2014pfa,Julie:2019sab}, giving rise to a ``fifth" force interaction between two such objects, along with scalar dipole radiation which increases the rate of inspiral in a binary~\cite{Takahiro}.

In this article, we forecast current and future constraints on the EdGB theory of gravity from proposed GW observations by testing the consistency between the expected inspiral and merger-ringdown signals~\cite{Ghosh_IMRcon,Ghosh_IMRcon2,Abbott_IMRcon,Abbott_IMRcon2}.
We consider EdGB corrections to not only the inspiral properties of a binary BH coalescence~\cite{Takahiro}, but also to the characteristic quasinormal modes (QNMs)~\cite{Blazquez-Salcedo:2016enn} and final properties of the post-merger BH~\cite{Ayzenberg:2014aka}.
To the best of our knowledge, the IMR consistency test has been put into context for an example modified theory of gravity for the first time, and can indeed be applied to other alternative theories of gravity, given the required ingredients.
See upcoming work by the same authors~\cite{Carson_BumpyQNM} for a similar analysis in the general, parameterized non-Kerr spacetimes.
Additionally, see another similar work by the same authors~\cite{Carson_QNMPRD} for a more thorough description of the analysis outlined in this article.

While the present analysis is not entirely robust, it is presented as a new alternative route to obtain order-of-magnitude estimates (or better in most scenarios) without the significant time concerns required with full numerical relativity (NR) solutions, which do not yet exist for most alternative theories of gravity\footnote{Additionally, such NR simulations face challenges such as a lack of numerically stable formulations which prevent them from being simulated with currently-known methods.}.
In particular, in the following analysis we only consider the leading-order post-Newtonian corrections to the waveform, we utilize a predictive Fisher analysis, we assume the QNMs are isospectral as they are in GR, and we neglect merger corrections to the merger-ringdown and only include the QNM ringdown corrections.
Such approximations lead the analysis to being less-robust, however it offers a new method to forecast estimated constraints on any given modified theory of gravity by taking into account additional pieces of information available to make the gravitational waveform closer to completion with a minimal degree of effort and computational time.


\section{Parameter Estimation}~\label{sec:PE}
In this section, we discuss the Fisher analysis~\cite{Poisson:Fisher,Berti:Fisher,Cutler:2007mi,Yagi:2009zm} techniques utilized in the main analysis to compute statistical and systematic uncertainties on template parameters $\theta^a$.
For loud enough events~\cite{Vallisneri:FisherSNR,Vallisneri:FisherSNR2}, a Fisher analysis approximation may be reliably used\footnote{A more comprehensive Bayesian analysis like that used by the LVC in e.g.~\cite{GW150914,GW170729} can be used to extract the true posterior probability distributions on source parameter}, yet is more computationally expensive. For loud enough events, the two have been shown to agree well. to estimate the likelihood function to provide approximate errors on recovered best-fit parameters $\theta^a$ from a given GW signal, with root-mean-square prior errors $\sigma_{\theta^a}^{(0)}$, and a waveform template $h$. We follow~\cite{Cutler:Fisher,Poisson:Fisher,Berti:Fisher} and assume knowledge of Gaussian prior probability distributions\footnote{A Bayesian analysis can utilize more natural prior probability distributions, such as uniform.}. 

\begin{figure}[htb]
\begin{center}
\includegraphics[width=.6\linewidth]{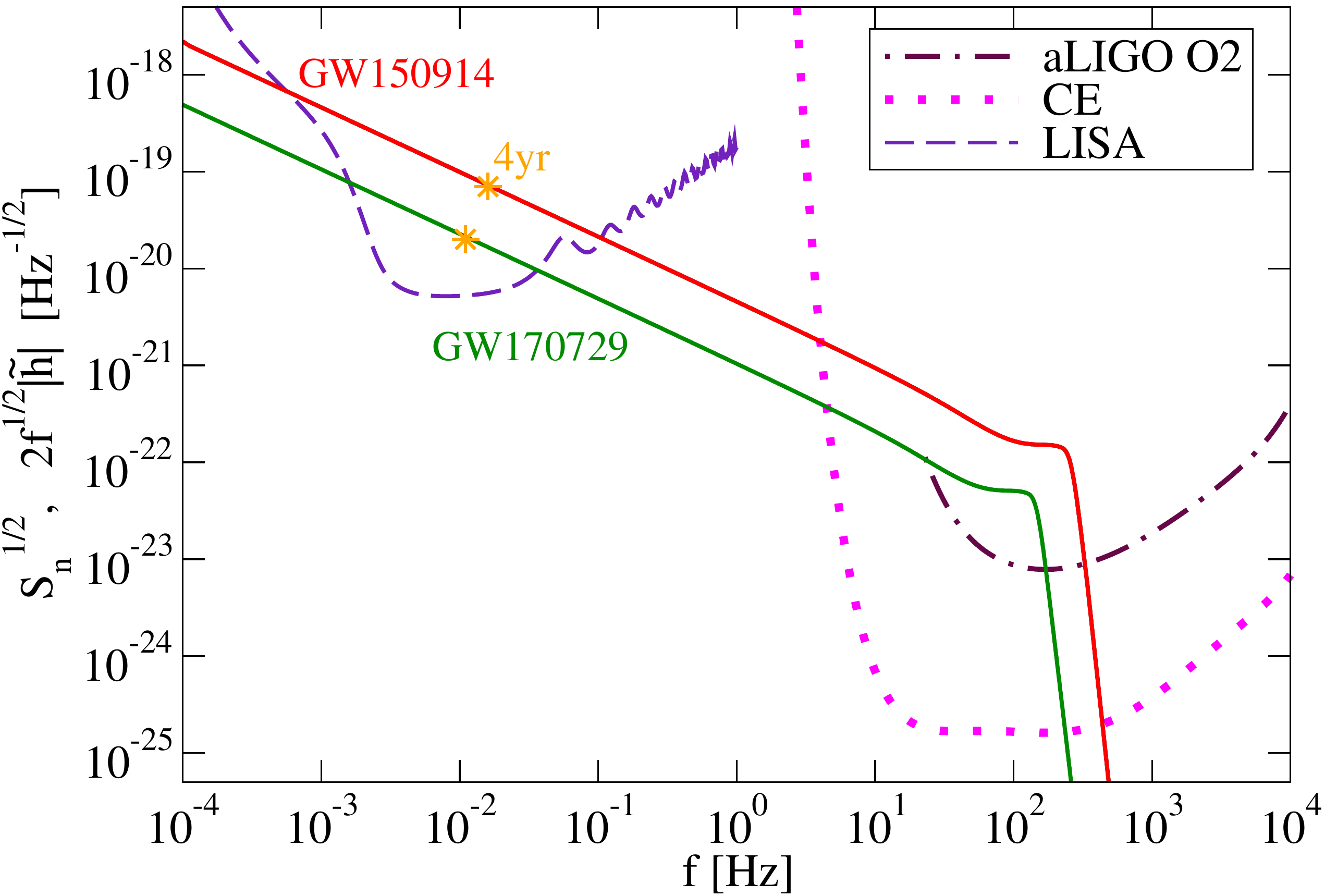}
\caption{
Sensitivities $\sqrt{S_n(f)}$ of the gravitational-wave interferometers aLIGO O2, CE, and LISA considered in this analysis.
We additionally display the characteristic amplitudes $2\sqrt{f}|\tilde h(f)|$ for GW events GW150914, and GW170729 with 4 years prior to merger displayed as orange stars.
}\label{fig:detectors}
\end{center}
\end{figure}

The statistical root-mean-square errors on parameters $\theta^a$ can be found to be
\begin{equation}
\Delta\theta^a=\sqrt{\tilde{\Gamma}_{ii}^{-1}},
\end{equation}
where $\tilde{\bm{\Gamma}}$ is the effective Fisher matrix $\tilde{\Gamma}_{ij}=\Gamma_{ij}+(\sigma_{\theta^a}^{(0)})^{-2}\delta_{ij}$, and the \emph{Fisher information matrix} can be given by
\begin{equation}
\Gamma_{ij}\equiv \left( \frac{\partial h}{\partial \theta^i} \Bigg| \frac{\partial h}{\partial \theta^j} \right).
\end{equation}
In the above expression, the notation $(a|b)$ represents the inner product weighted by the detector noise spectral density $S_n(f)$
\begin{equation}
(a|b)\equiv2\int^{f_\text{high}}_{f_\text{low}}\frac{\tilde{a}^*\tilde{b}+\tilde{b}^*\tilde{a}}{S_n(f)}df,
\end{equation}
where $f_\text{high,low}$ represent the detector-dependent high and low cutoff frequencies, as are tabulated and described in Table~\ref{tab:detectors}, taken and adapted from Ref.~\cite{Carson_multiBandPRD}.
In particular, we consider the ground-based detector aLIGO O2 (we used the fitted noise curve in App. C of~\cite{Yunes:2016jcc}), third generation detector Cosmic Explorer (CE)~\cite{Ap_Voyager_CE}, as well as future space-based detector LISA~\cite{LISA}, with detector sensitivities displayed in Fig.~\ref{fig:detectors}.
Finally, if one wishes to combine the detections from multiple detectors with Fisher matrices $\bm{\Gamma}^\text{A}$ and $\bm{\Gamma}^\text{B}$, the resulting effective Fisher matrix can be found to be
\begin{equation}
\tilde{\Gamma}^\text{tot}_{ij}=\Gamma^\text{A}_{ij}+\Gamma^\text{B}_{ij}+\frac{1}{(\sigma_{\theta^a}^{(0)})^2}\delta_{ij}.
\end{equation}

\begin{table*}
\centering
\resizebox{1.\textwidth}{!}{%
\begin{tabular}{c|c|c|c|c|c|c}
Detector & Location & GW150914 $f_{\text{low}}$ (Hz) & GW150914 $f_{\text{high}}$ (Hz) & GW150914 SNR & Arm length & interferometers\\
\hline
O2~\cite{aLIGO} & Ground & 23 & 400 & 24 & $4$ km & 1\\
CE~\cite{Ap_Voyager_CE} & Ground & 1 & 400 & $3.36 \times 10^3$ & $40$ km & 1\\
LISA~\cite{LISA} & Space & 0.02 & 1 & $9.30$ & 2.5 Gm & 2\\
\end{tabular}
}
\caption{
Tabulated information for the ground-based detectors O2 and CE and space-based detector LISA as considered in our analysis.
The lower ground-based and upper space-based frequency limits for GW150914-like events correspond to the detector limits $f_{\text{low-cut}}$ and $f_{\text{high-cut}}$, while the upper ground-based and lower space-based limits correspond to an arbitrary value such that the gravitational wave spectrum is sufficiently small compared to the detector sensitivity, and the GW frequency 4 years prior to merger~\cite{Carson_multiBandPRD}.
The GW150914 SNR is computed via $\sqrt{(h|h)}$.
}\label{tab:detectors}
\end{table*}

Additionally, following the analysis of Ref.~\cite{Cutler:2007mi}, one can estimate the ``theoretical", or systematic errors present in the extraction of template parameters $\theta^a$ due to mismodeling present in the template waveform. In particular, one can estimate systematic errors on $\theta^a$ by assuming use of the GR template, while EdGB gravity is in fact the correct theory in nature.
The theoretical errors can be computed as
\begin{equation}\label{eq:sys}
\Delta_\text{th}\theta^a \approx \Sigma^{ab} \left( \lbrack \Delta A+iA_\GR\Delta\Psi \rbrack e^{i\Psi_\GR} \Big| \partial_b \tilde{h}_\GR \right),
\end{equation}
where $\Sigma^{ab} = (\Gamma^{-1})^{ab}$ is the covariance matrix, a summation over $b$ is implied, and $\Delta A \equiv A_\GR-A_\EdGB$ and $\Delta\Psi \equiv \Psi_\GR-\Psi_\EdGB$ are the differences between the amplitude and phase in GR and EdGB gravity.
We note that the above expression for the systematic errors is most accurate when the difference between the GR and non-GR signals are small.
Thus for large enough values of EdGB coupling parameter the above approximation will become less accurate.
However, in the main analysis presented here, we impose the small coupling approximation which ensures small magnitudes of $\sqrt{\alpha_\EdGB}$.
The probability distributions in $(\Delta M_f/\bar{M}_f,\Delta \chi_f/\bar{\chi}_f)$ have included both statistical errors ($\sqrt{\bm{\Sigma}_{\I,\MR}}$) which determine their size, and systematic errors ($\bm{\Delta_\text{th}X}_{\I,\MR}$) which determine their offset from the GR predictions.

For the GR waveform, we utilize the non-precessing, sky-averaged \emph{IMRPhenomD} GR waveform presented through the NR fits of Refs.~\cite{PhenomDI,PhenomDII}.
The IMRPhenomD waveform is typically parameterized in terms of the $(\mathcal{M},\eta,\chi_s,\chi_a)$ mass and spin parameters, where $\chi_\text{s,a}\equiv (\chi_1 \pm \chi_s)/2$ are the symmetric and anti-symmetric aligned-spin combinations. However, in this investigation we re-parameterize it to instead include $(M_f,\eta,\chi_s,\chi_f)$.
This is accomplished by computing the expressions $\mathcal{M}(M_f,\eta,\chi_s,\chi_f,\zeta)$ and  $\chi_a(M_f,\eta,\chi_s,\chi_f,\zeta)$ from Eqs.~(12) and~(13) from the main text.
By re-parameterizing the template waveform like so, we can directly generate multi-dimensional posterior probability distributions with the final mass and spin $M_f$ and $\chi_f$, relevant in the analysis.
The resulting template waveform consists of
\begin{equation}
\theta^a=\left( \ln{\mathcal{A}},\phi_c,t_c,M_f,\eta,\chi_s,\chi_f,\zeta \right),
\end{equation}
where $\mathcal{A}\equiv\frac{\mathcal{M}_z^{5/6}}{\sqrt{30}\pi^{2/3}D_L}$ is a generalized amplitude with redshifted chirp mass $\mathcal{M}_z\equiv\mathcal{M}(1+z)$ and redshift $z$, $D_L$ is the luminosity distance, and $\phi_c$ and $t_c$ are the coalescence phase and time.
Additionally, we impose Gaussian prior distributions corresponding to $|\phi_c|\leq\pi$, $|\chi_s|\leq 1$, and $|\chi_f|\leq 1$ with 2-$\sigma$ errors.
Namely, the priors are imposed by taking the above upper bounds to be twice the standard deviation of a standard Gaussian distribution.

We then model an EdGB waveform by modifying the IMRPhenomD GR waveform in three ways. The first modification is in the inspiral portion, where we add the EdGB leading post-Newtonian correction as in Eq.~(2) of the main text.
The second modification is in the ringdown portion, where we modify the QNM ringdown and damping frequencies as in Eqs.~(3) and~(4) of the main text. The third modification is in the estimate of the final mass and spin, which is given in Eqs.~(12) and~(13) of the main text.
Finally, we compute the Fisher information matrix using the PhenomD GR waveform to approximate statistical errors on source parameters, and then using Eq.~\eqref{eq:sys} to estimate the systematic error ``shift'' one could expect to observe when detecting an EdGB signal described by our simple model.

We utilize fiducial values such that $\eta$ and $\chi_s$ correspond to the initial parameters of the GW event in question, $M_f$ and $\chi_f$ correspond to those predicted by Eqs.~(12) and~(13) in the main text, and $\phi_c=t_c=0$.
Finally, we allow the fiducial value of $\zeta$ to vary slowly as we proceed with the IMR consistency test with different magnitudes of EdGB coupling.


\section{Gravitational waveforms in Einstein-dilaton Gauss-Bonnet gravity}\label{sec:EdGBoverall}
EdGB gravity is an effective field theory in which a string-inspired ``dilaton" scalar field $\varphi$ is coupled to a quadratic curvature term~\cite{Kanti_EdGB,Maeda:2009uy,Sotiriou:2013qea} with coupling parameter $\alpha$. 
In particular, we consider the case where the scalar field is coupled linearly with curvature.~\cite{Kanti_EdGB}.
Scalar charges in EdGB gravity only anchor to BHs~\cite{Campbell:1991kz,Yunes:2011we,Takahiro,Sotiriou:2014pfa,Takahiro,Yagi:2015oca,Yagi:2011xp,Prabhu:2018aun}, and depend on their mass, spin, and $\alpha$.
For valid constraints on $\alpha$ to be placed, the small coupling approximation $\zeta \equiv \frac{16\pi\alpha^2}{M^4} \ll 1$ must be satisfied for binaries with total mass $M\equiv m_1+m_2$.
Current constraints on the theory have been found to be $\sqrt{\alpha} < 2$ km with GW observations using ppE corrections to the waveform~\cite{Yamada:2019zrb} and a low-mass X-ray binary~\cite{Yagi_EdGB} (see also~\cite{Kanti_EdGB,Pani_EdGB,Nair_dCSMap,Tahura:2019dgr}).
Let us describe below the various corrections to the gravitational waveform in EdGB gravity.

\subsection{Inspiral}

We begin with the inspiral portion of the waveform, which can be described in a parameterized post-Einsteinian (ppE) form~\cite{Yunes:2009ke}
\begin{equation}\label{eq:ppE}
\tilde{h}_{\text{ppE}}=A_{\GR}(f)(1+ \bar \alpha u^{-2})e^{i(\Psi_{\GR}(f)+\bar \beta u^{-7})}.
\end{equation}
Here $A_{\GR}$ and $\Psi_{\GR}$ are the GR amplitude and phase described by the \emph{IMRPhenomD} waveform~\cite{PhenomDII,PhenomDI}, $u=(\pi \mathcal{M} f)^{1/3}$ is the effective relative velocity of the inspiraling bodies with GW frequency $f$ and chirp mass $\mathcal{M}\equiv M \eta^{3/5}$ with symmetric mass ratio $\eta\equiv m_1 m_2/M^2$.
Further, $\bar \alpha$ $(\bar \beta)$ characterize the magnitude of the amplitude (phase) corrections given in~\cite{Takahiro,Tahura_GdotMap,Yagi_EdGBmap} as
\begin{equation}\label{eq:alpha_ppE}
\bar \alpha=-\frac{5}{192}\zeta\frac{(m_1^2 \tilde{s}_2-m_2^2\tilde{s}_1)^2}{M^4\eta^{18/5}}\,, \quad 
\bar \beta=-\frac{5}{7168}\zeta\frac{(m_1^2 \tilde{s}_2-m_2^2\tilde{s}_1)^2}{M^4\eta^{18/5}},
\end{equation}
where $\tilde s_A = 1-\chi_A^2/4 + \mathcal{O}(\chi_A^4)$ corresponds to the normalized scalar charge of the $A$-th body, and $\chi_A \equiv J/M^2$ are the dimensionless BH spin parameters with the angular momentum magnitude $J$.
We note that the following results and waveform corrections are carried out and valid in the small-spin approximation to quadratic order in BH spin for simplicity.

\subsection{Ringdown}

We next explain corrections to the ringdown portion of the waveform, which is characterized by the QNM ringdown and damping frequencies~\cite{Berti:2005ys,Berti:2009kk}.
We refer the readers to Ref.~\cite{Maselli:2019mjd} where similar corrections were made, and constraints with multiple GW events were quantified.
See also Refs.~\cite{McManus:2019ulj,Cardoso:2019mqo} where a general formalism to map ringdown corrections directly to specific theories of gravity was developed.
In this article, we consider corrections to the QNM frequencies to first order in the dimensionless coupling constant $\zeta$ as
\begin{align}
f_{\RD} &=f_{\RD,\GR}+\zeta f_{\RD,\zeta} + \mathcal{O}(\zeta^2),\\
f_{\text{damp}} &=f_{\text{damp},\GR}+\zeta f_{\text{damp},\zeta} + \mathcal{O}(\zeta^2),
\end{align}
where~$f_{\RD,\GR}$ and $f_{\text{damp},\GR}$ are the GR QNM frequency predictions~\cite{PhenomDI,PhenomDII}, while $f_{\RD,\zeta}$ and $f_{\text{damp},\zeta}$ are the EdGB corrections.
To derive such corrections $f_{\RD,\zeta}$ and $f_{\text{damp},\zeta}$ to first order in $\zeta$ and quadratic order in the final spin $\chi_f$, we use the results in Ref.~\cite{Blazquez-Salcedo:2016enn}\footnote{Reference~\cite{Blazquez-Salcedo:2016enn} follows a slightly different EdGB notation than considered here, beginning with the coupling parameter $\alpha$ in the action as well as their definition of $\zeta'$. 
The quantities can be mapped to our definitions by letting $\zeta' \rightarrow 4 \sqrt{\zeta}$.} to compute the complex QNM frequency up to quadratic order in spin $\chi_f$ of the remnant BH.
We take note of Ref.~\cite{Carson_QNMPRD} by the same authors, where this assumption was tested for accuracy.
In particular, it was found that by taking EdGB corrections to the waveform up to $\mathcal{O}(\chi^4)$ from Ref.~\cite{Maselli:2019mjd}, the resulting Fisher-estimated constraints only varied by $\sim1.5\%$ from the $\mathcal{O}(\chi^2)$ case, well within the accuracies of this analysis.
We consider only the leading, $\ell= m = 2$ axial QNMs. This is because the spinning components have only been computed for axial modes via null geodesics correspondence\footnote{See Refs.~\cite{Silva:2019scu,Glampedakis:2017dvb,Glampedakis:2019dqh} where the null geodesic correspondence was used to estimate corrections for rotating BHs.}~\cite{Yang:2012he}. One cannot use this correspondence to the polar modes since such modes are coupled to the scalar field perturbation. However, one expects the spin dependence on the polar mode to be comparable to that on the axial mode as an order of magnitude estimate~\cite{Blazquez-Salcedo:2016enn}. 
%
Finally, we find the EdGB corrections to the ringdown and damping frequencies as
%
\begin{align}
f_{\RD,\zeta}&= \frac{a_0(1+a_1\chi_f+a_2\chi_f^2)}{2\pi M_f} + \mathcal{O}(\chi_f^3),\label{eq:fRD}\\
f_{\text{damp},\zeta} &= \frac{b_0(1+b_1\chi_f+b_2\chi_f^2)}{2\pi M_f} + \mathcal{O}(\chi_f^3),
\label{eq:fdamp}
\end{align}
where $a_i$ and $b_i$ are presented in Table~\ref{tab:Mfchif}.
\begin{table}
\centering
\begin{tabular}{ c c c  }
$a_0$&$a_1$&$a_2$ \\
$-0.1874$ & $-0.6552$ & $-0.6385$ \\
\hline
$b_0$&$b_1$&$b_2$ \\
$-0.0622$ & $-0.1350$ & $-0.2251$ \\
\hline
\hline
$c_0$&$c_1$&$c_2$ \\
$\frac{43740-2233\sqrt{2}\eta^2}{262440\eta}$ & $\frac{50659\sqrt{3}\eta^2-116640\sqrt{6}}{12(2233\sqrt{2}\eta^2-43740)}$ & $\frac{1361569247\sqrt{2}\eta^2-1285956000}{264600(2233\sqrt{2}\eta^2-43740)}$ \\
\hline
$d_0$&$d_1$&$d_2$\\
$\frac{13571}{29160 \sqrt{3}}$&$\frac{75371}{40713}\sqrt{\frac{2}{3}}$&$\frac{58180627}{149620275}$
\end{tabular}
\caption{
Coefficients $a_i$ $b_i$, $c_i$, and $d_i$ required for the reconstruction of the EdGB corrections to the remnant BH QNM ringdown and damping frequencies $f_\RD$, $f_\text{damp}$, as well as the mass and spin $M_{f,\zeta}$ and $\chi_{f,\zeta}$ as found in Eqs.~\eqref{eq:fRD},~\eqref{eq:fdamp},~\eqref{eq:Mf} and~\eqref{eq:chif} respectively.
}\label{tab:Mfchif}
\end{table}

\subsection{Final mass and spin}

In addition to the inspiral and ringdown corrections discussed above, we also need to modify the predictions for the remnant BH's mass and spin under EdGB gravity.
Similar to the merger-ringdown corrections presented previously, we expand the final mass and spin of the remnant BH to first-order in $\zeta$ and second-order in $\chi_f$.
We take $M_{f,\GR}$ and $\chi_{f,\GR}$ to be the GR final mass and spin as presented by the NR fits of Ref.~\cite{PhenomDII}, while $M_{f,\zeta}$ and $\chi_{f,\zeta}$ are the first order EdGB corrections
\begin{align}
M_f &=M_{f,\GR}+\zeta M_{f,\zeta} + \mathcal{O}(\zeta^2),\label{eq:Mf} \\
\chi_f &=\chi_{f,\GR}+\zeta \chi_{f,\zeta} + \mathcal{O}(\zeta^2).\label{eq:chif}
\end{align}
In GR, the final mass and spin of the remnant BH can be estimated roughly from the initial masses $m_A$ and spins $\chi_A$ as the total orbital energy and angular momentum of a test particle with mass $\mu$ orbiting around a BH with mass $M_f (\sim M)$ and $\chi_f$ at the innermost stable circular orbit (ISCO)~\cite{Barausse:2009uz}, or
\begin{eqnarray}
\label{eq:E_Mf}
\mu \left[1-E_\text{orb}(M_f,\chi_f,r_\ISCO)\right]&=&M-M_f, \\
\mu L_\text{orb}(M,\chi_f,r_\ISCO)&=&M(M \chi_f-a_s-\delta_m a_a)
\end{eqnarray}
Here $a_{s,a}\equiv (m_1\chi_1 \pm m_2\chi_2)/2$, $\mu$ is the reduced mass, $\delta_m \equiv (m_1-m_2)/M$ is the weighted mass difference, while $E_\mathrm{orb}$ and $L_\mathrm{orb}$ are the specific orbital energy and orbital angular momentum respectively (can be found in Eqs.~(63)--(68) of~\cite{Ayzenberg:2014aka}), and $r_\ISCO$ is the location of the ISCO.
We assume that the same picture still holds in theories beyond GR~\cite{Yunes_ModifiedPhysics}. 
Additionally, in EdGB gravity there is a scalar interaction energy realized between the orbiting scalarized bodies~\cite{Stein:2013wza} which contributes to the radiated mass, so Eq.~\eqref{eq:E_Mf} is then modified to
\begin{equation}
\mu \left[1-E_\text{orb}(M_f,\chi_f,r_\ISCO)-E_\text{scalar}(\mu,M,\chi_f,r_\ISCO,\zeta) \right]=M-M_f,
\end{equation}
with~\cite{Stein:2013wza}
\begin{equation}
E_\text{scalar}(\mu,M,\chi_f,r_\ISCO,\zeta)=\frac{\zeta}{\eta ^2}\left(1-\frac{\chi_f^2}{4}\right) \frac{M}{r},
\end{equation}
corresponding to the specific scalar interaction energy between the particle (with mass $\mu$ and zero spin) and the central BH (with mass $M_f$ and spin $\chi_f$).
Having these expressions at hand, one can estimate the EdGB corrections to these quantities as
\begin{align}
M_{f,\zeta}&= M c_0 \left( 1+c_1 \chi_f+c_2\chi_f^2 \right)  + \mathcal{O}\left(\chi_f^3\right), \label{eq:Mf}\\
\chi_{f,\zeta} &= -d_0\eta \left(1 + d_1 \chi_{f,\GR} + d_2 \chi_{f,\GR}^2\right) + \mathcal{O}\left(\chi_{f,\GR}^3\right),\label{eq:chif}
\end{align}
where $c_i$ and $d_i$ are presented in Table~\ref{tab:Mfchif}.
Observe that the above expressions themselves depend on the remnant BH spin in GR ($\chi_{f,\GR}$), found in Ref.~\cite{PhenomDII}.
We also note that the above expression for $M_f$ depends on the solution for $\chi_f$. 
$M_f$ is then expanded once again after $\chi_f$ has been substituted in to quadratic order in spin.


\section{Theory agnostic tests of GR}\label{sec:testsOfGR}In this section, we present the theory-agnostic IMR consistency test of GR~\cite{Ghosh_IMRcon,Ghosh_IMRcon2,Abbott_IMRcon,Abbott_IMRcon2,Zack:Proceedings}.
In accordance with the no-hair theorem, the post-coalescense BH can be described by only two parameters: the mass $M_f(m_1,m_2,\chi_1,\chi_2)$ and spin $\chi_f(m_1,m_2,\chi_1,\chi_2)$, which can be estimated with the NR fits of Ref.~\cite{PhenomDI}.
Assuming GR were to be the true theory of gravity found in nature, such final parameters may be uniquely predicted by each of the inspiral GW signal (I, $f<f_\ISCO=(6^{3/2}\pi M)^{-1}$) and the merger-ringdown GW signal\footnote{Due to the absence of NR modeling of non-GR waveforms, merger corrections are absent from this analysis. Simplified ringdown corrections are included in the consistency test, as well as corrections to the final mass and spin.} (MR, $f>f_\ISCO$).
On the other hand, if such signals were to disagree on their final parameter predictions, an emergent modified theory of gravity (such as EdGB) may be present.

The IMR consistency test can be performed as follows. 
First, we generate the two-dimensional posterior probability distributions $P_{\I,\MR}(M_f,\chi_f)$ in the $M_f-\chi_f$ plane from each portion of the waveform described above.
Such posterior distributions are described as a two-dimensional Gaussian probability distribution function with root-mean-square errors estimated via a Fisher-based analysis, as described below.
Next, we measure the agreement between these two signals by transforming the I and MR probability distributions into $\Delta M_f/\bar{M}_f$ and $\Delta\chi_f/\bar{\chi}_f$~\cite{Ghosh_2017}, where $\Delta M_f\equiv M_f^\I-M_f^\MR$ and $\Delta \chi_f \equiv \chi_f^\I-\chi_f^\MR$ describe the differences between the inspiral and merger-ringdown GR predictions of the final mass and spin, and $\bar{M}_f\equiv\frac{1}{2}(M_f^\I+M_f^\MR)$ and $\bar{\chi}_f\equiv\frac{1}{2}(\chi_f^\I+\chi_f^\MR)$ describe the averages between the two.
Finally, the consistency of the probability distribution in the $(\Delta M_f/\bar{M}_f,\Delta \chi_f/\bar{\chi}_f)$ plane with the GR value of $(0,0)$ determines the agreement with GR, while any statistically significant deviations may be interpreted as evidence towards emergent non-GR effects present within the observed signal.
See similar analyses by the same authors~\cite{Carson_multiBandPRD,Zack:Proceedings,Carson_QNMPRD} for a more detailed description of the IMR consistency test.

All detected GW signals to date have been found to be consistent with GR~\cite{Abbott_IMRcon,Abbott_IMRcon2,Ghosh_IMRcon,Ghosh_IMRcon2,Ghosh_2017}.
Reference~\cite{Ghosh_IMRcon2} phenomenologically introduced a non-GR correction at second post-Newtonian order in the gravitational wave energy flux and studied the IMR consistency test, though the reference did not include corrections to the QNM ringdown spectrum for simplicity.

\begin{figure}[htb]
\begin{center}
\includegraphics[width=0.6\textwidth]{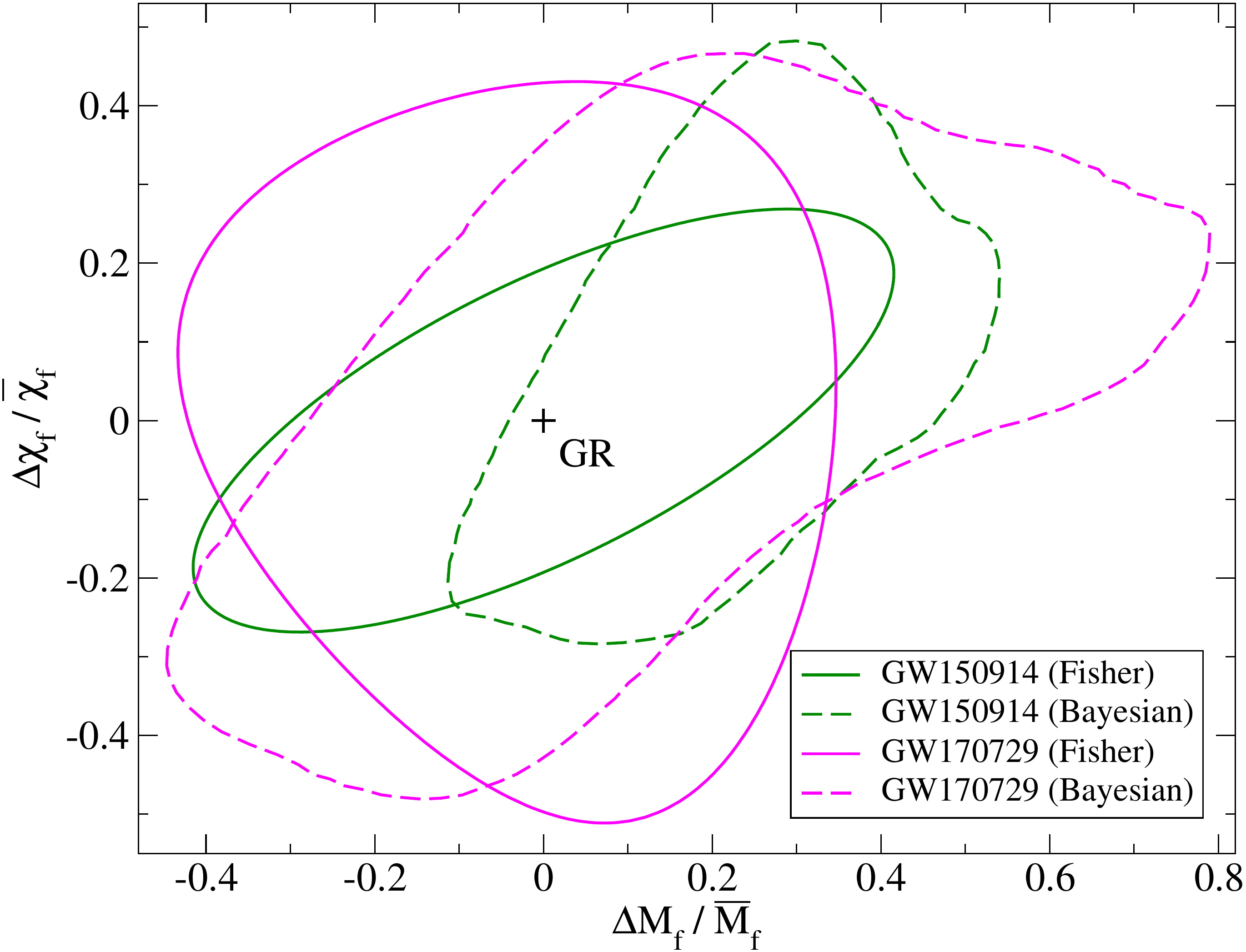}
\caption{
Comparison between the transformed posterior probability distributions in the IMR consistency test for both the Fisher analysis method (solid) used here, and the Bayesian one (dashed) done by the LVC in~\cite{Abbott_IMRcon}.
We display the results for both GW events GW150914 (green) and GW170729 (magenta) considered in this analysis.
We observe that in both cases, the total enclosed areas of the probability distributions agree between the Fisher and Bayesian analyses to within 10\% accuracy, confirming that the former can capture some qualitative features of the latter and thus is reliable as an order-of-magnitude estimate, at least for the magnitude of statistical uncertainties.
We do note that in the case of GW170729 the Fisher analysis distribution does not quite represent the \textit{correlations} observed in the more comprehensive Bayesian analysis: something we expect to improve with future high-SNR observations.
}\label{fig:FisherBayesian}
\end{center}
\end{figure}

In this analysis, we follow closely along with the investigation found in Refs.~\cite{Carson_multiBandPRD,Zack:Proceedings}.
Namely, we utilize a Fisher-analysis-based technique, rather than the typical Bayesian analysis found in~\cite{Ghosh_IMRcon,Ghosh_IMRcon2,Abbott_IMRcon,Abbott_IMRcon2}.
While the latter analysis is more accurate and allows one to calculate the location of the posterior probability distributions, the former analysis allows one to approximate the \emph{size} of such distributions, under the assumption of loud GW events, and Gaussian noise, prior distributions, and posterior distributions.
See Sec.~\ref{sec:PE} for a brief description of our Fisher analysis method.
While this is not particularly useful for current GW events, it is of high value when predicting the non-GR resolving power of future detectors.
Combined with the \emph{theoretical} (systematic) uncertainty ``shifts" $\bm{\Delta_\text{th}X}_{\I,\MR}\equiv(\Delta_\text{th} M_f,\Delta_\text{th}\chi_f)$ described in Ref.~\cite{Cutler:2007mi} and Sec.~\ref{sec:PE},
the final probability distributions in the $M_f-\chi_f$ plane are taken to be Gaussian
\begin{align}
P_{\I,\MR}&\equiv \frac{1}{2\pi\sqrt{|\bm{\Sigma}_{\I,\MR}|}}\exp \left\lbrack -\frac{1}{2} \left(  \bm{X} - \bm{X}^\GR_{\I,\MR} -\bm{\Delta_\text{th}X}_{\I,\MR} \right)^\text{T} \right. \nonumber \\
&\hspace{4cm} \left. \times \bm{\Sigma}_{\I,\MR}^{-1}\left( \bm{X} - \bm{X}^\GR_{\I,\MR} -\bm{\Delta_\text{th}X}_{\I,\MR} \right) \right\rbrack,\label{eq:pdf}
\end{align}
where $\bm{\Sigma}_{\I,\MR}$ represents the covariance matrix, $\bm{X}\equiv(M_f,\chi_f)$ contains the final state variables, and $\bm{X}^\GR_{\I,\MR}$ contains their GR predictions from the inspiral and merger-ringdown portions respectively.
See Fig.~\ref{fig:FisherBayesian} for a comparison between the Fisher analysis method considered here, and the Bayesian done by the LVC in~\cite{Abbott_IMRcon}.
We note that for both GW150914 and GW170729, the total enclosed areas of the Fisher and Bayesian probability distributions agree with each other to within 10\% accuracy.
This confirms the validity of the Fisher analysis method considered in this paper as a qualitative estimate, something which we expect to improve considerably for future observations with increased SNRs.
We do note, however, that while the total areas agree well (indicative of the total statistical uncertainties), the direction of correlations for the case of GW170729 do not agree particularly well.
We expect this to improve as well for future high-SNR events.

We apply this method to test EdGB gravity as follows. 
We choose the template waveform to be the IMRPhenomD waveform in GR, while we inject a signal in EdGB gravity by implementing the EdGB corrections to the inspiral, ringdown and final mass/spin of the IMRPhenomD waveform given by Eqs.~\eqref{eq:ppE}--\eqref{eq:chif}.
We increase the fiducial value of $\zeta$ from 0 ($\bm{\Delta_\text{th}X}_{\I,\MR}=0$), until finally the GR prediction of $(\Delta M_f/\bar{M}_f,\Delta \chi_f/\bar{\chi}_f)|_\GR=(0,0)$ falls outside of the 90\% confidence region (i.e. the systematic uncertainties are larger than the statistical errors).
This indicates the magnitude of $\zeta$ required to observe non-GR effects in the waveform.
We note that in the following presented analysis, the explicit role of the theoretical error found in~\cite{Cutler:2007mi} is different than that used in Ref.~\cite{Cutler:2007mi} and a similar multi-band analysis paper~\cite{Glampedakis:2019dqh}.
In both of the above references, the authors describe such theoretical error as a source of theoretical uncertainty due to mismodeling of the waveform.
On the other hand, in this analysis the theoretical errors are used to simulate the shift that best-fit parameters $M_f$ and $\chi_f$ would experience given EdGB corrections were present in the true signal while the GR waveform has been used for the data analysis.
Such shifts in best-fit parameters are then directly compared to the parameter covariances found with the Fisher information matrix.
We believe this is the first analysis where the IMR consistency tests have been applied to a concrete non-GR theory, where both inspiral and ringdown corrections are consistently included.


\section{Results}\label{sec:results}
Now let us discuss the resulting detectability of EdGB effects using the IMR consistency tests of GR, using the process outlined in the prior section.
This is done by injecting varying magnitudes of EdGB effects into the waveform until the IMR consistency test is failed.

\begin{figure*}[htb]
\begin{center}
\includegraphics[width=.46\textwidth]{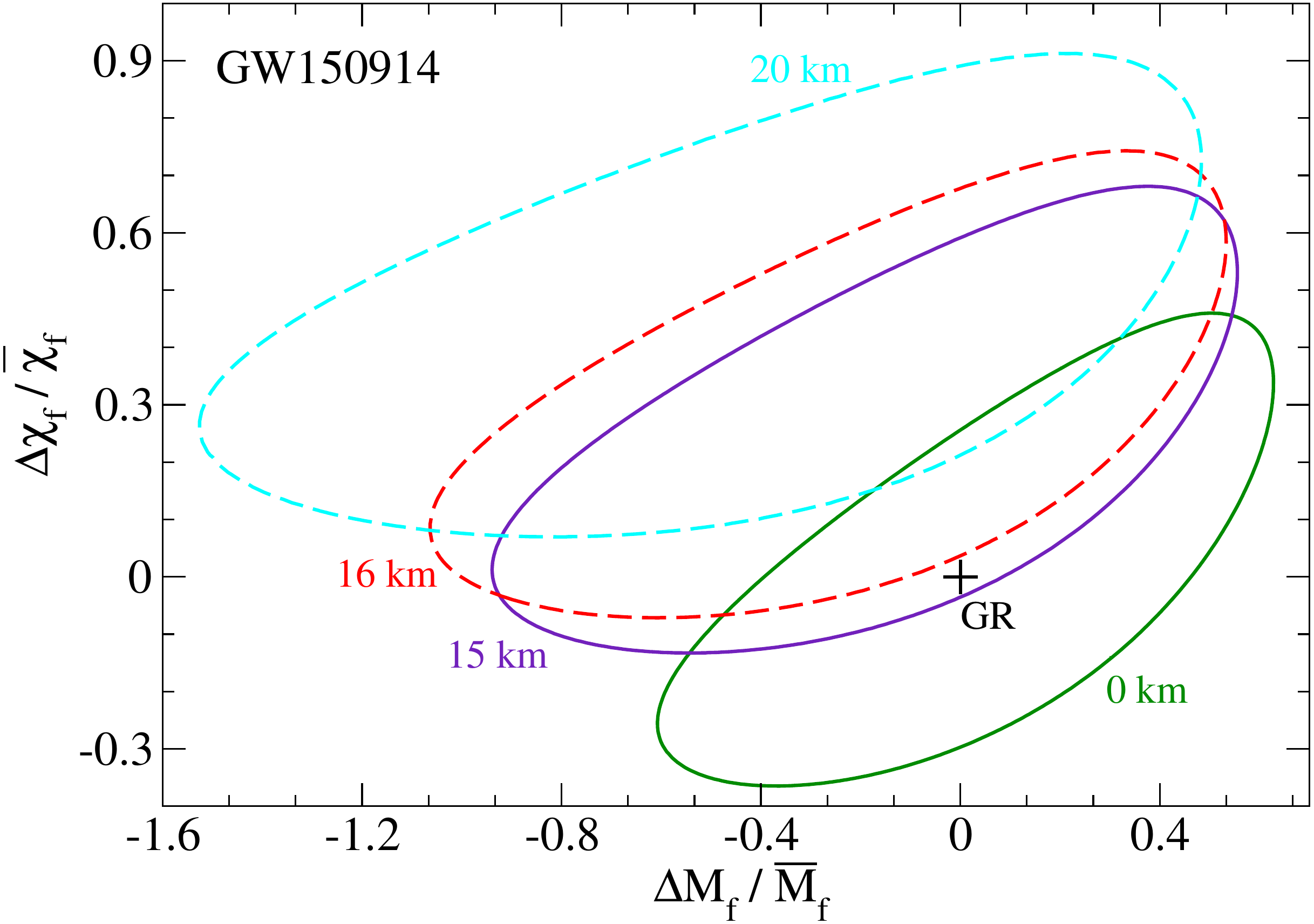}
\includegraphics[width=.46\textwidth]{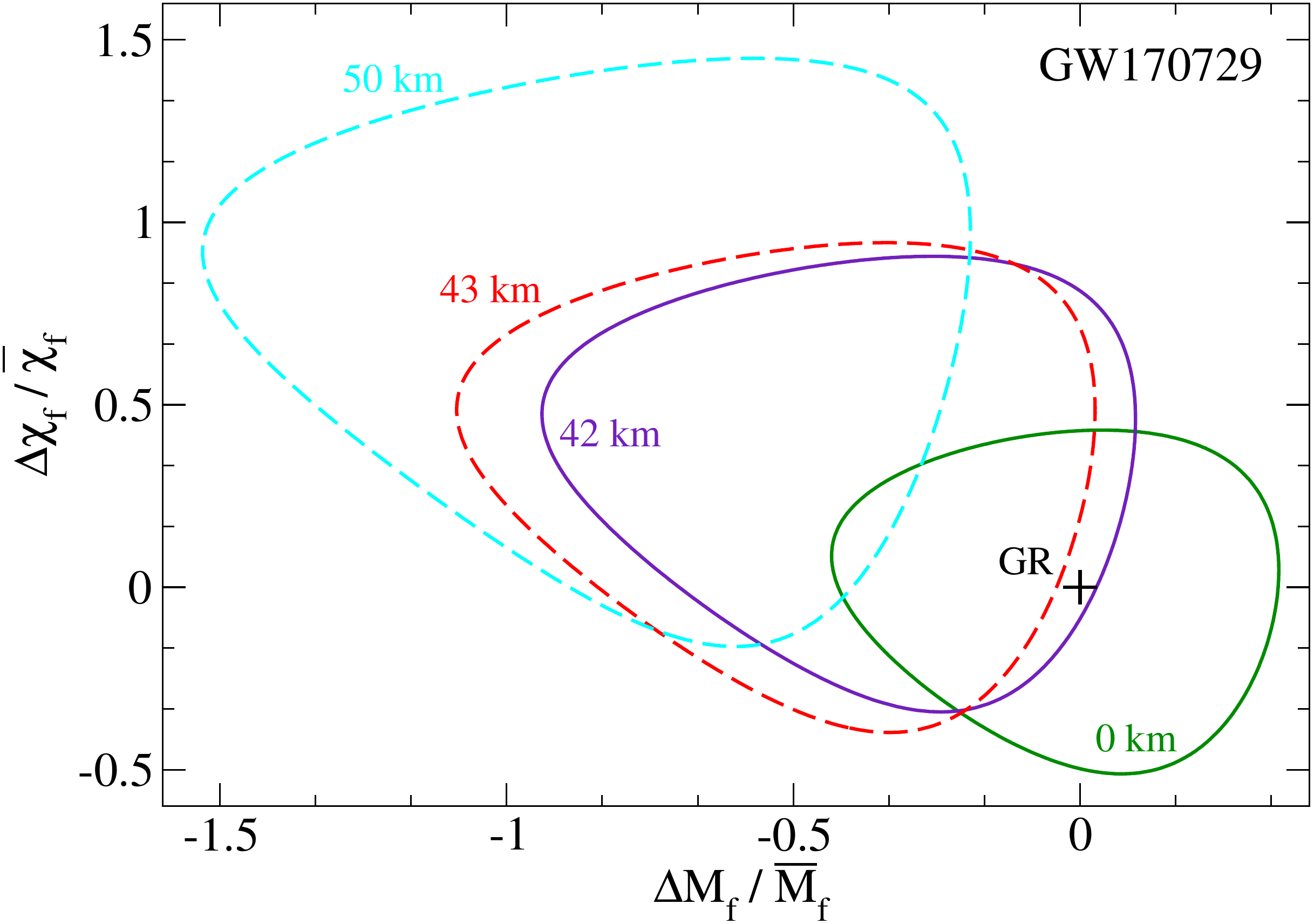}
\caption{
(left) IMR consistency test performed under EdGB gravity for GW150914 with a corresponding waveform generated via IMRPhenomD with the O2 detector.
Displayed is the 90\% confidence region of the transformed probability distribution in the $\Delta M_f/\bar{M}_f-\Delta \chi_f/\bar{\chi}_f$ plane, with the GR value of $(0,0)$.
The analysis is repeated for various fiducial values of $\sqrt{\alpha}$.
(right) Same as the left panel but for the more massive GW event GW170729 with $(M_f,\chi_f)_\GR=(80.3\text{ M}_\odot,0.81)$.
}\label{fig:IMRD_O2}
\end{center}
\end{figure*}

Let us first discuss the current prospects of observing EdGB effects upon the detection of binary BH merger events by the LIGO O2~\cite{Yunes:2016jcc} (see App. C) detector.
The left panel of Fig.~\ref{fig:IMRD_O2} present the results of the test for GW150914 with $\sqrt{\alpha}=(0\text{ km},15\text{ km},16\text{ km}, 20\text{ km})$.
Such a waveform was generated with the PhenomD model assuming BH masses and spins of $(m_1,m_2,\chi_1,\chi_2)=(38.9 M_\odot,31.6M_\odot,0.32,-0.44)$, with a luminosity distance scaled to a signal-to-noise (SNR) ratio of $25.1$.
The above masses and spins were obtained from the median values of each distribution as reported in~\cite{GW150914}, and the alignment of the spins were chosen to be in agreement with the median value of effective spin $\chi_\text{eff}$ as reported by the same reference.
We observe that, at the 90\% confidence interval, EdGB effects can be observed for $\sqrt{\alpha}>15$ km, much larger than the current constraint of $2$ km~\cite{Kanti_EdGB,Pani_EdGB,Yagi_EdGB,Nair_dCSMap,Yamada:2019zrb,Tahura:2019dgr}.
Therefore, we confirm that the current LVC infrastructure is unable to detect EdGB effects based on the existing observational constraints of $\sqrt{\alpha}<2$ km.
Similarly, we repeat the process for the more massive event GW170729 in the right panel of Fig.~\ref{fig:IMRD_O2}, observing how contributions from the merger-ringdown signal are much more significant in this scenario, with large uncertainties now present in the inspiral signal instead, resulting in EdGB detectability of $\sqrt{\alpha}>42$ km.
This waveform was generated with BH masses and spins of $(m_1,m_2,\chi_1,\chi_2)=(50.6M_\odot,34.3M_\odot,0.60,-0.57)$ with an SNR of $10.8$.
Similar to before, the masses and spins were chosen from the median values reported in~\cite{GW170729}, with the spin alignments chosen to be in agreement with $\chi_\text{eff}$.

\begin{figure}[htb]
\begin{center}
\includegraphics[width=.45\textwidth]{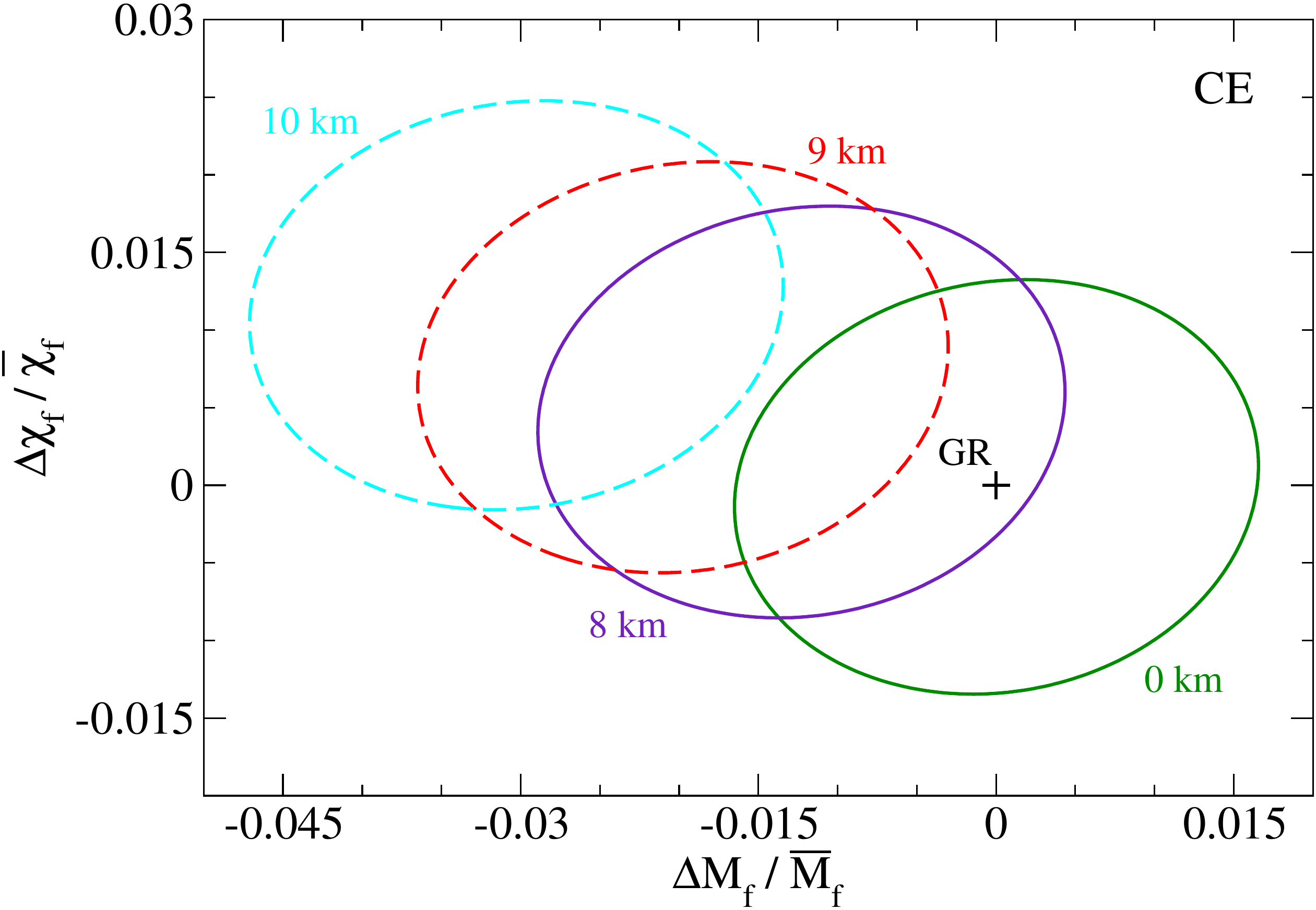}
\includegraphics[width=.45\textwidth]{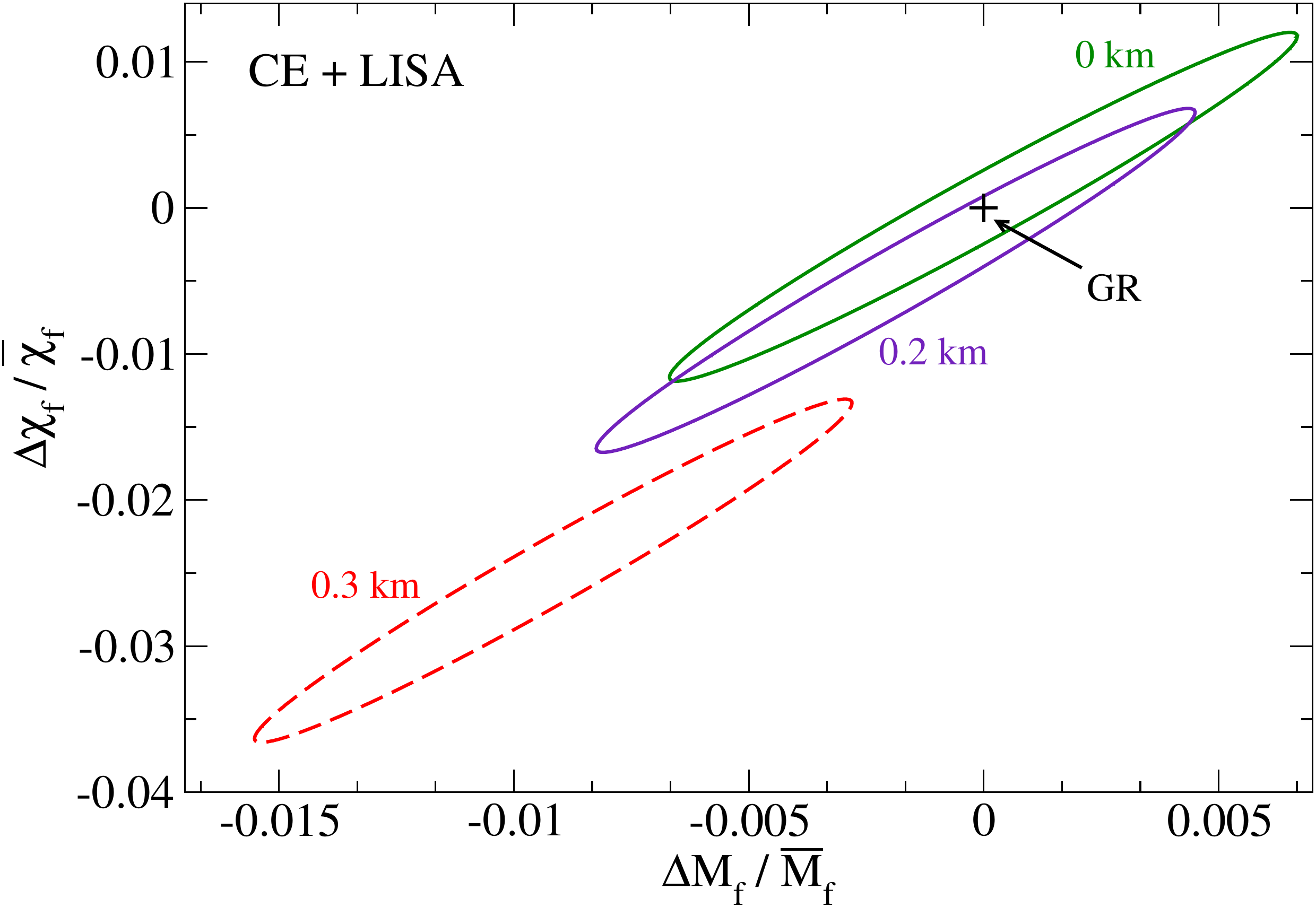}
\caption{
Similar to Fig.~\ref{fig:IMRD_O2} but with the CE detector (left), and the multi-band observation between LISA and CE (right).
}\label{fig:IMRD_future}
\end{center}
\end{figure}

Next we consider the future prospects of observing such effects in the waveform with third-generation ground-based detectors.
The left panel of Fig.~\ref{fig:IMRD_future} shows the resulting probability distributions in $(\Delta M_f/\bar{M}_f,\Delta \chi_f/\bar{\chi}_f)$ found with the Cosmic Explorer~\cite{Ap_Voyager_CE} (CE) observations of GW150914-like events, with $\sqrt{\alpha}=(0\text{ km},8\text{ km},9\text{ km},10\text{ km})$.
We see that with CE, EdGB effects can be determined to a 90\% confidence interval for $\sqrt{\alpha}>8$ km, still above the current constraint of $\sqrt{\alpha}<2$ km.

By noting that a majority of both the statistical (size of the contours) and systematic uncertainties (shift of the contour centers) come from the inspiral signal, we consider a multiband observation by combining CE with the space-based detector LISA~\cite{LISA} to further probe the inspiral event.
The right panel of Fig.~\ref{fig:IMRD_future} shows the resulting probability distributions in $(\Delta M_f/\bar{M}_f,\Delta \chi_f/\bar{\chi}_f)$ observed by multiband observations with $\sqrt{\alpha}=(0\text{ km},0.2\text{ km},0.3\text{ km})$.
Here we see that multiband detections can probe EdGB effects with magnitudes of $\sqrt{\alpha}>0.2$ km, an order-of-magnitude smaller than the current constraint of $2$ km.
Thus, if non-GR effects such as EdGB are indeed present in nature with $0.2\text{ km}<\sqrt{\alpha}<2$ km, multiband detections between CE and LISA can uncover them to the 90\% confidence interval. On the other hand, if one does not find deviations from GR, one would be able to place bounds on EdGB gravity that are stronger than current bounds by an order of magnitude.
The projected bounds with future detectors presented here using IMR consistency tests are comparable to those found with parameterized tests of GR~\cite{Zack:Proceedings,Carson_multiBandPRL,Carson_multiBandPRD}.

In the above analysis we have shown the considerable increase one might gain by introducing milli-Hz era GW detectors such as LISA (or e.g. TianQin~\cite{TianQin}) to the ground-based observations with third-generation detector CE.
We note that such constraints can also be expected to improve considerably upon the additional observation from deci-Hz detectors such as (B-)DECIGO~\cite{B-DECIGO,DECIGO}.
As found in Ref.~\cite{Carson_multiBandPRD} by the same authors, multi-band observations with CE+DECIGO considerably outperformed those with any other space-based detector such as LISA or TianQin.
In particular, Table~I of~\cite{Carson_multiBandPRD} shows such bounds to improve anywhere from a factor of two to two orders-of-magnitude when introducing multi-band observations with DECIGO.
For the specific theory of EdGB, constraints on $\sqrt{\alpha_\EdGB}$ were shown to improve by nearly a factor of three.


\section{Discussion}\label{sec:conclusion}
In this article, we chose EdGB gravity as an example non-GR theory to study the power of IMR consistency tests, though the formalism that we developed here can easily be applied to other theories if all of the ingredients are available. 
For example, dynamical Chern-Simons gravity is another theory beyond GR that breaks parity and is motivated from string theory, loop quantum gravity and effective field theory for inflation~\cite{Jackiw:2003pm,Alexander_cs,Nishizawa:2018srh}. 
Leading post-Newtonian corrections to the inspiral waveform have been derived in~\cite{Yagi:2012vf,Nair_dCSMap,Tahura:2019dgr}, while the scalar interaction energy and corrections to the specific orbital energy, angular momentum and the location of the ISCO have been computed in~\cite{Stein:2013wza,Yagi_dCS,Yagi:2012vf}. The QNM ringdown spectrum for non-spinning BHs in such a theory has been studied in~\cite{Yunes:2007ss,Cardoso:2009pk,Molina:2010fb}. Therefore, once the spin corrections to the BH ringdown in this theory is available, one can repeat the analysis here to investigate how accurately one can probe dynamical Chern-Simons gravity with the IMR consistency test.

Although we have taken into account the known EdGB corrections to the waveform to date as much as possible, there are some other modifications that we have left out. 
Below, we list some of the caveats in our analysis presented in this article:
\begin{itemize}
\item We only include leading-order post-Newtonian terms in the waveform, while a more advanced analysis could include higher order corrections. Though such corrections do not seem to be important in certain scalar-tensor theories~\cite{Yunes_ModifiedPhysics} (see App. B).
\item In GR, axial and polar QNMs are identical (isospectral), while such isospectrality is broken in EdGB~\cite{Blazquez-Salcedo:2016enn}. Thus, the ringdown portion of the waveform may be more complicated than that for GR.
\item Our estimate for the mass and spin of the BH remnant in EdGB gravity is based on the picture verified in GR, though this needs to be justified once NR simulations of binary black hole mergers are available in such a theory~\cite{Witek:2018dmd}.
\item We did not include corrections during the merger phase of the waveform. Again, it is likely that one needs to wait for NR simulations to realize how the corrections enter in this phase.
\end{itemize}
Having said this, we believe our calculations should be valid as an order of magnitude estimate. One reason to support this point is because corrections to the waveform enter linearly in $\zeta \propto \alpha^2$. Thus, even if our estimates are off by an order of magnitude in $\zeta$, bounds on $\sqrt{\alpha}$ are affected only by a factor of $\sim 10^{1/4} \sim 2$.

For the purposes and scope of this investigation, the Fisher analysis has been used to predict valid order-of-magnitude constraints on the EdGB theory of gravity.
As thoroughly discussed in Ref.~\cite{Yunes_ModifiedPhysics} as well as~\cite{Carson_multiBandPRD,Zack:Proceedings}, for large enough SNR the results approximate well a Bayesian analysis. In the former reference, the Fisher-estimated non-GR parameter $\bar{\beta}$ in the inspiral agreed with its Bayesian counterpart to within $\sim40\%$ at $-1$PN for GW150914-like events with an SNR of 25. This corresponds to only a $\sim10\%$ difference on the coupling parameter $\sqrt{\alpha}$ in EdGB gravity. Regarding the latter, the 90\% credible contours in the final mass-spin plane obtained with Fisher and Bayesian analyses agreed with an error of 20\% for GW150914. Such agreements only strengthen considerably for the future detectors considered in this analysis.

In the above investigation, we utilized an approximate Fisher analysis based approach to predict posterior probability distributions on BH source parameters by assuming fiducial values given by the median values reported by the LVC.
A more comprehensive analysis would instead make use of the maximal likelihood values of such source parameters obtained directly from posterior probability distributions.
For highly skewed posterior probability distributions, the difference between the two values could potentially be significant.
Albeit, we find this point to be beyond the scope of this analysis, which is provided as a first step approximation to test GR with order-of-magnitude estimates on source parameters.
We leave this point, as well as a full Bayesian analysis to future work.

If a non-GR effect is observed, how can one potentially infer whether it originated from one non-GR theory or another? Given that the inspiral-merger-ringdown tests discussed here were originally designed to test the consistency of GR, a different test would be more appropriate to address the above question. For example, one could directly try to measure the leading corrections to the inspiral and ringdown frequency independently and check for the consistency between the two quantities within a given non-GR theory. In the case of EdGB gravity, one can eliminate $\zeta$ from the two to find such a relation, which is unique to the theory.

\emph{Additional note}--- While this work was nearing completion, the first numerical relativity simulations for binary black hole mergers in EdGB gravity including corrections to the metric perturbations were carried out in Ref.~\cite{Okounkova:2020rqw,Okounkova:2019zjf} (see also Ref.~\cite{Okounkova:2019dfo} for a related work in dynamical Chern-Simons gravity).
We plan to compare those waveforms against the ones presented here to check the validity of the latter and aim to more correctly account for the merger-ringdown corrections, which we leave for future work.


\section{Acknowledgments}\label{acknowledgments}
We thank Maria Okounkova and Leo Stein for valuable comments.
Z.C. and K.Y. acknowledge support from NSF Award PHY-1806776 and the Ed Owens Fund. K.Y. would like to also acknowledge support by a Sloan Foundation Research Fellowship, the COST Action GWverse CA16104 and JSPS KAKENHI Grants No. JP17H06358.

\section*{References}

\bibliographystyle{iopart-num}
\bibliography{Zack}

\newcommand{\noop}[1]{}
\providecommand{\newblock}{}
\begin{thebibliography}{10}
\expandafter\ifx\csname url\endcsname\relax
  \def\url#1{{\tt #1}}\fi
\expandafter\ifx\csname urlprefix\endcsname\relax\def\urlprefix{URL }\fi
\providecommand{\eprint}[2][]{\url{#2}}

\bibitem{GW150914}
Abbott B~P {\em et~al.\/} (LIGO Scientific, Virgo) 2016 {\em Phys. Rev.
  Lett.\/} {\bf 116} 241102 (\textit{Preprint} \eprint{1602.03840})

\bibitem{GW_Catalogue}
Abbott B~P {\em et~al.\/} (LIGO Scientific, Virgo) 2018  (\textit{Preprint}
  \eprint{1811.12907})

\bibitem{Abbott_IMRcon}
Abbott B~P {\em et~al.\/} (LIGO Scientific, Virgo) 2019  (\textit{Preprint}
  \eprint{1903.04467})

\bibitem{Will_SEP}
Will C~M 2014 {\em Living Rev. Rel.\/} {\bf 17} 4 (\textit{Preprint}
  \eprint{1403.7377})

\bibitem{TheVirgo:2014hva}
Acernese F {\em et~al.\/} (VIRGO) 2015 {\em Class. Quant. Grav.\/} {\bf 32}
  024001 (\textit{Preprint} \eprint{1408.3978})

\bibitem{TheLIGOScientific:2014jea}
Aasi J {\em et~al.\/} (LIGO Scientific) 2015 {\em Class. Quant. Grav.\/} {\bf
  32} 074001 (\textit{Preprint} \eprint{1411.4547})

\bibitem{advancedLIGO}
and J~Aasi {\em et~al.\/} 2015 {\em Classical and Quantum Gravity\/} {\bf 32}
  074001 \urlprefix\url{https://doi.org/10.1088%2F0264-9381%2F32%2F7%2F074001}

\bibitem{Ap_Voyager_CE}
Ligo-t1400316-v4: Instrument science white paper
  \url{https://dcc.ligo.org/ligo-T1400316/public}
  \urlprefix\url{https://dcc.ligo.org/ligo-T1400316/public}

\bibitem{ET}
The {ET} project website \url{http://www.et-gw.eu/}
  \urlprefix\url{http://www.et-gw.eu/}

\bibitem{LISA}
Robson T, Cornish N and Liu C 2019 {\em Class. Quant. Grav.\/} {\bf 36} 105011
  (\textit{Preprint} \eprint{1803.01944})

\bibitem{B-DECIGO}
Isoyama S, Nakano H and Nakamura T 2018 {\em PTEP\/} {\bf 2018} 073E01
  (\textit{Preprint} \eprint{1802.06977})

\bibitem{DECIGO}
Yagi K and Seto N 2011 {\em Phys. Rev.\/} {\bf D83} 044011 [Erratum: Phys.
  Rev.D95,no.10,109901(2017)] (\textit{Preprint} \eprint{1101.3940})

\bibitem{TianQin}
Shi C, Bao J, Wang H, Zhang J~d, Hu Y, Sesana A, Barausse E, Mei J and Luo J
  2019  (\textit{Preprint} \eprint{1902.08922})

\bibitem{Will_SolarSystemTest}
Will C~M 2014 {\em Living Reviews in Relativity\/} {\bf 17} 4 ISSN 1433-8351
  \urlprefix\url{https://doi.org/10.12942/lrr-2014-4}

\bibitem{Stairs_BinaryPulsarTest}
Stairs I~H 2003 {\em Living Rev. Rel.\/} {\bf 6} 5 (\textit{Preprint}
  \eprint{astro-ph/0307536})

\bibitem{Wex_BinaryPulsarTest}
Wex N 2014  (\textit{Preprint} \eprint{1402.5594})

\bibitem{Ferreira_CosmologyTest}
Ferreira P~G 2019  (\textit{Preprint} \eprint{1902.10503})

\bibitem{Clifton_CosmologyTest}
Clifton T, Ferreira P~G, Padilla A and Skordis C 2012 {\em Phys. Rept.\/} {\bf
  513} 1--189 (\textit{Preprint} \eprint{1106.2476})

\bibitem{Joyce_CosmologyTest}
Joyce A, Jain B, Khoury J and Trodden M 2015 {\em Phys. Rept.\/} {\bf 568}
  1--98 (\textit{Preprint} \eprint{1407.0059})

\bibitem{Koyama_CosmologyTest}
Koyama K 2016 {\em Rept. Prog. Phys.\/} {\bf 79} 046902 (\textit{Preprint}
  \eprint{1504.04623})

\bibitem{Salvatelli_CosmologyTest}
Salvatelli V, Piazza F and Marinoni C 2016 {\em JCAP\/} {\bf 1609} 027
  (\textit{Preprint} \eprint{1602.08283})

\bibitem{Berti_ModifiedReviewLarge}
Berti E {\em et~al.\/} 2015 {\em Class. Quant. Grav.\/} {\bf 32} 243001
  (\textit{Preprint} \eprint{1501.07274})

\bibitem{Yunes_ModifiedPhysics}
Yunes N, Yagi K and Pretorius F 2016 {\em Phys. Rev. D\/} {\bf 94}(8) 084002
  \urlprefix\url{https://link.aps.org/doi/10.1103/PhysRevD.94.084002}

\bibitem{Jain:2010ka}
Jain B and Khoury J 2010 {\em Annals Phys.\/} {\bf 325} 1479--1516
  (\textit{Preprint} \eprint{1004.3294})

\bibitem{Salvatelli:2016mgy}
Salvatelli V, Piazza F and Marinoni C 2016 {\em JCAP\/} {\bf 1609} 027
  (\textit{Preprint} \eprint{1602.08283})

\bibitem{Koyama:2015vza}
Koyama K 2016 {\em Rept. Prog. Phys.\/} {\bf 79} 046902 (\textit{Preprint}
  \eprint{1504.04623})

\bibitem{Joyce:2014kja}
Joyce A, Jain B, Khoury J and Trodden M 2015 {\em Phys. Rept.\/} {\bf 568}
  1--98 (\textit{Preprint} \eprint{1407.0059})

\bibitem{Famaey:2011kh}
Famaey B and McGaugh S 2012 {\em Living Rev. Rel.\/} {\bf 15} 10
  (\textit{Preprint} \eprint{1112.3960})

\bibitem{Milgrom:DarkMatter}
{Milgrom} M 1983 {\em Astrophys. J.\/} {\bf 270} 365--370

\bibitem{Milgrom:2008rv}
Milgrom M 2008  (\textit{Preprint} \eprint{0801.3133})

\bibitem{Clifton:2011jh}
Clifton T, Ferreira P~G, Padilla A and Skordis C 2012 {\em Phys. Rept.\/} {\bf
  513} 1--189 (\textit{Preprint} \eprint{1106.2476})

\bibitem{Kanti_EdGB}
Kanti P, Mavromatos N~E, Rizos J, Tamvakis K and Winstanley E 1996 {\em Phys.
  Rev.\/} {\bf D54} 5049--5058 (\textit{Preprint} \eprint{hep-th/9511071})

\bibitem{Maeda:2009uy}
Maeda K~i, Ohta N and Sasagawa Y 2009 {\em Phys. Rev.\/} {\bf D80} 104032
  (\textit{Preprint} \eprint{0908.4151})

\bibitem{Sotiriou:2013qea}
Sotiriou T~P and Zhou S~Y 2014 {\em Phys. Rev. Lett.\/} {\bf 112} 251102
  (\textit{Preprint} \eprint{1312.3622})

\bibitem{Campbell:1991kz}
Campbell B~A, Kaloper N and Olive K~A 1992 {\em Phys. Lett.\/} {\bf B285}
  199--205

\bibitem{Yunes:2011we}
Yunes N and Stein L~C 2011 {\em Phys. Rev.\/} {\bf D83} 104002
  (\textit{Preprint} \eprint{1101.2921})

\bibitem{Takahiro}
Yagi K and Tanaka T 2010 {\em Prog. Theor. Phys.\/} {\bf 123} 1069--1078
  (\textit{Preprint} \eprint{0908.3283})

\bibitem{Sotiriou:2014pfa}
Sotiriou T~P and Zhou S~Y 2014 {\em Phys. Rev.\/} {\bf D90} 124063
  (\textit{Preprint} \eprint{1408.1698})

\bibitem{Julie:2019sab}
Julie F~L and Berti E 2019  (\textit{Preprint} \eprint{1909.05258})

\bibitem{Ghosh_IMRcon}
Ghosh A {\em et~al.\/} 2016 {\em Phys. Rev.\/} {\bf D94} 021101
  (\textit{Preprint} \eprint{1602.02453})

\bibitem{Ghosh_IMRcon2}
Ghosh A, Johnson-Mcdaniel N~K, Ghosh A, Mishra C~K, Ajith P, Del~Pozzo W, Berry
  C~P~L, Nielsen A~B and London L 2018 {\em Class. Quant. Grav.\/} {\bf 35}
  014002 (\textit{Preprint} \eprint{1704.06784})

\bibitem{Abbott_IMRcon2}
Abbott B~P {\em et~al.\/} (LIGO Scientific, Virgo) 2016 {\em Phys. Rev.
  Lett.\/} {\bf 116} 221101 [Erratum: Phys. Rev. Lett.121,no.12,129902(2018)]
  (\textit{Preprint} \eprint{1602.03841})

\bibitem{Blazquez-Salcedo:2016enn}
Bl\'{a}zquez-Salcedo J~L, Macedo C~F~B, Cardoso V, Ferrari V, Gualtieri L, Khoo
  F~S, Kunz J and Pani P 2016 {\em Phys. Rev.\/} {\bf D94} 104024
  (\textit{Preprint} \eprint{1609.01286})

\bibitem{Ayzenberg:2014aka}
Ayzenberg D and Yunes N 2014 {\em Phys. Rev.\/} {\bf D90} 044066 [Erratum:
  Phys. Rev.D91,no.6,069905(2015)] (\textit{Preprint} \eprint{1405.2133})

\bibitem{Carson_BumpyQNM}
Carson Z and Yagi K 2020 {\em Phys. Rev. D\/} {\bf 101}(8) 084050
  \urlprefix\url{https://link.aps.org/doi/10.1103/PhysRevD.101.084050}

\bibitem{Carson_QNMPRD}
Carson Z and Yagi K 2020 {\em Phys. Rev. D\/} {\bf 101}(10) 104030
  (\textit{Preprint} \eprint{2003.00286})
  \urlprefix\url{https://link.aps.org/doi/10.1103/PhysRevD.101.104030}

\bibitem{Poisson:Fisher}
Poisson E and Will C~M 1995 {\em Phys. Rev. D\/} {\bf 52}(2) 848--855
  \urlprefix\url{https://link.aps.org/doi/10.1103/PhysRevD.52.848}

\bibitem{Berti:Fisher}
Berti E, Buonanno A and Will C~M 2005 {\em Phys. Rev.\/} {\bf D71} 084025
  (\textit{Preprint} \eprint{gr-qc/0411129})

\bibitem{Cutler:2007mi}
Cutler C and Vallisneri M 2007 {\em Phys. Rev.\/} {\bf D76} 104018
  (\textit{Preprint} \eprint{0707.2982})

\bibitem{Yagi:2009zm}
Yagi K and Tanaka T 2010 {\em Phys. Rev.\/} {\bf D81} 064008 [Erratum: Phys.
  Rev.D81,109902(2010)] (\textit{Preprint} \eprint{0906.4269})

\bibitem{Vallisneri:FisherSNR}
Vallisneri M 2011 {\em Phys. Rev. Lett.\/} {\bf 107}(19) 191104
  \urlprefix\url{https://link.aps.org/doi/10.1103/PhysRevLett.107.191104}

\bibitem{Vallisneri:FisherSNR2}
Vallisneri M 2008 {\em Phys. Rev. D\/} {\bf 77}(4) 042001
  \urlprefix\url{https://link.aps.org/doi/10.1103/PhysRevD.77.042001}

\bibitem{GW170729}
Chatziioannou K {\em et~al.\/} 2019  (\textit{Preprint} \eprint{1903.06742})

\bibitem{Cutler:Fisher}
Cutler C and Flanagan E~E 1994 {\em Phys. Rev. D\/} {\bf 49}(6) 2658--2697
  \urlprefix\url{https://link.aps.org/doi/10.1103/PhysRevD.49.2658}

\bibitem{Carson_multiBandPRD}
Carson Z and Yagi K 2020 {\em Phys. Rev. D\/} {\bf 101}(4) 044047
  (\textit{Preprint} \eprint{1911.05258})
  \urlprefix\url{https://link.aps.org/doi/10.1103/PhysRevD.101.044047}

\bibitem{Yunes:2016jcc}
Yunes N, Yagi K and Pretorius F 2016 {\em Phys. Rev.\/} {\bf D94} 084002
  (\textit{Preprint} \eprint{1603.08955})

\bibitem{aLIGO}
Advanced {LIGO} \url{https://www.advancedligo.mit.edu/}
  \urlprefix\url{https://www.advancedligo.mit,.edu/}

\bibitem{PhenomDI}
Khan S, Husa S, Hannam M, Ohme F, P\"urrer M, Forteza X~J and Boh\'e A 2016
  {\em Phys. Rev. D\/} {\bf 93}(4) 044007
  \urlprefix\url{https://link.aps.org/doi/10.1103/PhysRevD.93.044007}

\bibitem{PhenomDII}
Husa S, Khan S, Hannam M, P\"urrer M, Ohme F, Forteza X~J and Boh\'e A 2016
  {\em Phys. Rev. D\/} {\bf 93}(4) 044006
  \urlprefix\url{https://link.aps.org/doi/10.1103/PhysRevD.93.044006}

\bibitem{Yagi:2015oca}
Yagi K, Stein L~C and Yunes N 2016 {\em Phys. Rev.\/} {\bf D93} 024010
  (\textit{Preprint} \eprint{1510.02152})

\bibitem{Yagi:2011xp}
Yagi K, Stein L~C, Yunes N and Tanaka T 2012 {\em Phys. Rev.\/} {\bf D85}
  064022 [Erratum: Phys. Rev.D93,no.2,029902(2016)] (\textit{Preprint}
  \eprint{1110.5950})

\bibitem{Prabhu:2018aun}
Prabhu K and Stein L~C 2018 {\em Phys. Rev.\/} {\bf D98} 021503
  (\textit{Preprint} \eprint{1805.02668})

\bibitem{Yamada:2019zrb}
Yamada K, Narikawa T and Tanaka T 2019  (\textit{Preprint} \eprint{1905.11859})

\bibitem{Yagi_EdGB}
Yagi K 2012 {\em Phys. Rev.\/} {\bf D86} 081504 (\textit{Preprint}
  \eprint{1204.4524})

\bibitem{Pani_EdGB}
Pani P and Cardoso V 2009 {\em Phys. Rev.\/} {\bf D79} 084031
  (\textit{Preprint} \eprint{0902.1569})

\bibitem{Nair_dCSMap}
Nair R, Perkins S, Silva H~O and Yunes N 2019  (\textit{Preprint}
  \eprint{1905.00870})

\bibitem{Tahura:2019dgr}
Tahura S, Yagi K and Carson Z 2019 {\em Phys. Rev.\/} {\bf D100} 104001
  (\textit{Preprint} \eprint{1907.10059})

\bibitem{Yunes:2009ke}
Yunes N and Pretorius F 2009 {\em Phys. Rev.\/} {\bf D80} 122003
  (\textit{Preprint} \eprint{0909.3328})

\bibitem{Tahura_GdotMap}
Tahura S and Yagi K 2018 {\em Phys. Rev.\/} {\bf D98} 084042 (\textit{Preprint}
  \eprint{1809.00259})

\bibitem{Yagi_EdGBmap}
Yagi K, Stein L~C, Yunes N and Tanaka T 2012 {\em Phys. Rev.\/} {\bf D85}
  064022 [Erratum: Phys. Rev.D93,no.2,029902(2016)] (\textit{Preprint}
  \eprint{1110.5950})

\bibitem{Berti:2005ys}
Berti E, Cardoso V and Will C~M 2006 {\em Phys. Rev.\/} {\bf D73} 064030
  (\textit{Preprint} \eprint{gr-qc/0512160})

\bibitem{Berti:2009kk}
Berti E, Cardoso V and Starinets A~O 2009 {\em Class. Quant. Grav.\/} {\bf 26}
  163001 (\textit{Preprint} \eprint{0905.2975})

\bibitem{Maselli:2019mjd}
Maselli A, Pani P, Gualtieri L and Berti E 2020 {\em Phys. Rev.\/} {\bf D101}
  024043 (\textit{Preprint} \eprint{1910.12893})

\bibitem{McManus:2019ulj}
McManus R, Berti E, Macedo C~F~B, Kimura M, Maselli A and Cardoso V 2019 {\em
  Phys. Rev.\/} {\bf D100} 044061 (\textit{Preprint} \eprint{1906.05155})

\bibitem{Cardoso:2019mqo}
Cardoso V, Kimura M, Maselli A, Berti E, Macedo C~F~B and McManus R 2019 {\em
  Phys. Rev.\/} {\bf D99} 104077 (\textit{Preprint} \eprint{1901.01265})

\bibitem{Silva:2019scu}
Silva H~O and Glampedakis K 2019  (\textit{Preprint} \eprint{1912.09286})

\bibitem{Glampedakis:2017dvb}
Glampedakis K, Pappas G, Silva H~O and Berti E 2017 {\em Phys. Rev.\/} {\bf
  D96} 064054 (\textit{Preprint} \eprint{1706.07658})

\bibitem{Glampedakis:2019dqh}
Glampedakis K and Silva H~O 2019 {\em Phys. Rev.\/} {\bf D100} 044040
  (\textit{Preprint} \eprint{1906.05455})

\bibitem{Yang:2012he}
Yang H, Nichols D~A, Zhang F, Zimmerman A, Zhang Z and Chen Y 2012 {\em Phys.
  Rev.\/} {\bf D86} 104006 (\textit{Preprint} \eprint{1207.4253})

\bibitem{Barausse:2009uz}
Barausse E and Rezzolla L 2009 {\em Astrophys. J.\/} {\bf 704} L40--L44
  (\textit{Preprint} \eprint{0904.2577})

\bibitem{Stein:2013wza}
Stein L~C and Yagi K 2014 {\em Phys. Rev.\/} {\bf D89} 044026
  (\textit{Preprint} \eprint{1310.6743})

\bibitem{Zack:Proceedings}
Carson Z and Yagi K 2019 {Parameterized and Consistency Tests of Gravity with
  Gravitational Waves: Current and Future} {\em {Proceedings, Recent Progress
  in Relativistic Astrophysics: Shanghai, China, May 6-8, 2019}\/}
  (\textit{Preprint} \eprint{1908.07103})

\bibitem{Ghosh_2017}
Ghosh A, Johnson-McDaniel N~K, Ghosh A, Mishra C~K, Ajith P, Pozzo W~D, Berry
  C~P~L, Nielsen A~B and London L 2017 {\em Classical and Quantum Gravity\/}
  {\bf 35} 014002 \urlprefix\url{https://doi.org/10.1088%2F1361-6382%2Faa972e}

\bibitem{Carson_multiBandPRL}
Carson Z and Yagi K 2020 {\em Class. Quant. Grav.\/} {\bf 37} 02LT01
  (\textit{Preprint} \eprint{1905.13155})

\bibitem{Jackiw:2003pm}
Jackiw R and Pi S~Y 2003 {\em Phys. Rev.\/} {\bf D68} 104012 (\textit{Preprint}
  \eprint{gr-qc/0308071})

\bibitem{Alexander_cs}
Alexander S and Yunes N 2009 {\em Phys. Rept.\/} {\bf 480} 1--55
  (\textit{Preprint} \eprint{0907.2562})

\bibitem{Nishizawa:2018srh}
Nishizawa A and Kobayashi T 2018 {\em Phys. Rev.\/} {\bf D98} 124018
  (\textit{Preprint} \eprint{1809.00815})

\bibitem{Yagi:2012vf}
Yagi K, Yunes N and Tanaka T 2012 {\em Phys. Rev. Lett.\/} {\bf 109} 251105
  [Erratum: Phys. Rev. Lett.116,no.16,169902(2016)] (\textit{Preprint}
  \eprint{1208.5102})

\bibitem{Yagi_dCS}
Yagi K, Yunes N and Tanaka T 2012 {\em Phys. Rev.\/} {\bf D86} 044037 [Erratum:
  Phys. Rev.D89,049902(2014)] (\textit{Preprint} \eprint{1206.6130})

\bibitem{Yunes:2007ss}
Yunes N and Sopuerta C~F 2008 {\em Phys. Rev.\/} {\bf D77} 064007
  (\textit{Preprint} \eprint{0712.1028})

\bibitem{Cardoso:2009pk}
Cardoso V and Gualtieri L 2009 {\em Phys. Rev.\/} {\bf D80} 064008 [Erratum:
  Phys. Rev.D81,089903(2010)] (\textit{Preprint} \eprint{0907.5008})

\bibitem{Molina:2010fb}
Molina C, Pani P, Cardoso V and Gualtieri L 2010 {\em Phys. Rev.\/} {\bf D81}
  124021 (\textit{Preprint} \eprint{1004.4007})

\bibitem{Witek:2018dmd}
Witek H, Gualtieri L, Pani P and Sotiriou T~P 2019 {\em Phys. Rev.\/} {\bf D99}
  064035 (\textit{Preprint} \eprint{1810.05177})

\bibitem{Okounkova:2020rqw}
Okounkova M 2020  (\textit{Preprint} \eprint{2001.03571})

\bibitem{Okounkova:2019zjf}
Okounkova M, Stein L~C, Moxon J, Scheel M~A and Teukolsky S~A 2019
  (\textit{Preprint} \eprint{1911.02588})

\bibitem{Okounkova:2019dfo}
Okounkova M, Stein L~C, Scheel M~A and Teukolsky S~A 2019 {\em Phys. Rev.\/}
  {\bf D100} 104026 (\textit{Preprint} \eprint{1906.08789})

\end{thebibliography}
\end{document}